\documentclass[reqno]{article}
\usepackage{alltt}
\usepackage{graphicx}
\begin{document}

\title{Quantum Geometry of the Dynamical Space-time}
\author{P. Leifer$^1$}
\date{Bar-Ilan University, Ramat-Gan, Israel}
\maketitle \footnotetext[1]{On leave from Crimea State Engineering
and Pedagogical University, Simferopol, Crimea, Ukraine}
\begin{abstract}
Quantum theory of field (extended) objects without a priori
space-time geometry has been represented. Intrinsic coordinates in
the tangent fibre bundle over complex projective Hilbert state space
$CP(N-1)$ are used instead of space-time coordinates. The fate of
quantum system modeled by the generalized coherent states is rooted
in this manifold. Dynamical (state-dependent) space-time arises only
at the stage of the quantum ``yes/no" measurement. The quantum
measurement of the gauge ``field shell'' of the generalized coherent
state is described in terms of the affine parallel transport of the
local dynamical variables in $CP(N-1)$.
\end{abstract}
\vskip 0.1cm
\section{Introduction}
Strings, membranes are most popular now models of extended quantum
objects. They were evoked mostly in order to avoid divergences in
quantum field theory. These speculative models give rise for
interesting cosmological constructions that are very far from the
mature paradigm of our Universe. But I have a doubt that we have now
reliable tools for the discussions of global space-time structure
(quantum cosmology) since behind achievements in quantum physics we
have fundamental unsolved problems in the foundations of quantum
mechanics (QM), quantum field theory (QFT), etc.

I would like to show in this work that there are different kind of
extended quantum objects which I called ``field shell" of
generalized coherent states (GCS's) and that in the framework of
this model, the local structure of the quantum (or dynamical)
space-time has a natural Minkowskian 4D form.

Fast progress achieved in QFT due to efforts of Dirac, Fermi,
Jordan, Wigner and other physicists/mathematecans led to jump over
gap between one-body problem, which mostly is the academic one, to
many-body problem, which is, of course, more practical. However the
practical success of the QM and QFT cannot hide the principle
difficulties in the one-body quantum problem. I will recall shortly
these difficulties: the deep disagreement between both special and
general relativity and quantum theory \cite{Penrose}, in particular:
delocalization of the relativistic wave-function \cite{Heg};
divergences problem in QFT; the problem of the mass spectrum, etc.
The necessity of the quantum geometry as an universal tool evoked to
remove difficulties in foundations of quantum theory has been
already discussed \cite{Ash}. One of the main obstacle on this way
is just the problem of quantum interactions and their applications
to QM measurement problem as the one of the most important in the
modern physics \cite{Penrose}.

Macroscopic objects are typically well individualizable because of
weak interaction and this is the base for our abstraction to
material points and further to mathematical points which mutually
differ only by space-time coordinates. In quantum area we generally
loss this possibility. But of course the conservation of some
fundamental dynamical variables does exist. This conservation is
related to some internal symmetry of a quantum system and it gives a
mechanism of identification of some quantum particle in the
framework of one-body problem. One-body measurement (besides bubble
chamber measurements) is rarely achievable. The most close to
one-body quantum measurement is measurement on coherent beams. I
would like to show in present work that the polarization
measurements of coherent photons described from the point of view of
the breakdown of $U(2)$ symmetry \cite{Le1}, may be generalized to
the measurement of arbitrary Hermitian dynamical variable of a
``N-level" system as a  breakdown of the $G=SU(N)$ symmetry.

\section{The Action Quantization}
Schr\"odinger sharply denied the existence of so-called ``quantum
jumps" during the process of emission/absorption of the quants of
energy (particles) \cite{Schr1,Schr2}. Leaving the question about
the nature of quantum particles outside of consideration, he thought
about these processes as a resonance of the de Broglie waves that
phenomenologically may look like ``jumps" between two ``energy
levels". The second quantization method formally avoids these
questions but there are at least two reasons for its modification:

{\it First.} In the second quantization method one has formally
given particles whose properties are defined by some commutation
relations between creation-annihilation operators. Note, that the
commutation relations are only the simplest consequence of the
curvature of the dynamical group manifold in the vicinity of the
group's unit (in algebra). Dynamical processes require, however,
finite group transformations and, hence, the global group structure.
The main my technical idea is to use vector fields over a group
manifold instead of Dirac's abstract q-numbers. This scheme
therefore seeks the dynamical nature of the creation and
annihilation processes of quantum particles.

{\it Second.} The quantum particles (energy bundles) should
gravitate. Hence, strictly speaking, their behavior cannot be
described as a linear superposition. Therefore the ordinary second
quantization method (creation-annihilation of free particles) is
merely a good approximate scheme due to the weakness of gravity.
Thereby the creation and annihilation of particles are time
consuming dynamical non-linear processes. So, linear operators of
creation and annihilation (in Dirac sense) do exist as approximate
quantities.

I discuss here a modification of the ``second quantization''
procedure. One may image some one-dimension ``chain" of the {\it
action states of ``elementary quantum motion" (EAS) $|\hbar a>$ with
entire number $a$ of the action quanta}. These $a,b,c,...$ takes the
place of the ``principle quantum number" serving as discrete indices
$0 \leq a,b,c... <~ \infty$. Since the action does not create
gravity in itself, it is possible to create the linear superposition
of $|\hbar a>=(a!)^{-1/2} ({\hat \eta^+})^a|\hbar 0>$ constituting
$SU(\infty)$ multiplete of the Planck's action quanta operator
$\hat{S}=\hbar {\hat \eta^+} {\hat \eta}$ with the spectrum
$S_a=\hbar a$ in the separable Hilbert space $\cal{H}$. Therefore,
we shall primarily quantize the action, and not the energy. The
relative (local) vacuum of some problem is not necessarily the state
with minimal energy, it is a state with an extremal of some action
functional.

The space-time representation of these states and their coherent
superposition $|F>=\sum_{a=0}^{\infty}f^a |\hbar a>$ is postponed on
the dynamical stage as it is described below. The superposition
principle being applied universally to the EQS $|\hbar a>$, i.e. not
only for the description of processes of photon reflection on a
mirror, refraction in a glass, etc., \cite{Feynman}, but to quantum
particles, leads to conclusion that quantum particles arose as some
{\it dynamical processes}. In other words the Schr\"odinger's idea
of shaping stable wave packet (as the model of material point of the
classical oscillator) \cite{Schr3} from oscillators wave functions
will applied now to the shaping of {\it the action amplitudes of an
arbitrary quantum system}. Their evolution is defined pure
geometrically by some internal unitary gauge ``field shell" of the
quantum particle. It is close to the old idea of the ``field of
unitary spin" splitting the hadrons super-multiplete. The stationary
processes are represented by stable particles and quasi-stationary
processes are represented by unstable resonances.

POSTULATE 1.

\noindent {\it I assume that there are elementary quantum states
$|\hbar a>, a=0,1,...$ of an abstract Planck oscillator whose states
correspond to the quantum motions with given number of Planck action
quanta}.

We shall construct non-linear field equations describing energy
(frequency) distribution in the ``chain" of the EAS's $|\hbar a>$,
whose soliton-like solution provides the quantization of the
dynamical variables. Quantum ``particles'', and, hence, their
numbers arise as some countable solutions of non-linear wave
equations. In order to establish acceptable field equations which
are capable of intrinsically describing all possible degrees of
freedom under intensive interaction we construct a {\it universal
ambient Hilbert state space} $\cal{H}$. We will use {\it the
universality of the action} whose variation is capable of generating
any relevant dynamical variable.

Generally the coherent superposition
\begin{eqnarray}
|F>=\sum_{a=0}^{\infty} f^a| \hbar a>,
\end{eqnarray}
may represent of a ground state or a ``vacuum" of some quantum
system with the action operator
\begin{eqnarray}
\hat{S}=\hbar A({\hat \eta^+} {\hat \eta}).
\end{eqnarray}
Then one can define the action functional
\begin{eqnarray}
S[|F>]=\frac{<F|\hat{S}|F>}{<F|F>},
\end{eqnarray}
which has the eigen-value $S[|\hbar a>]=\hbar a$ on the eigen-vector
$|\hbar a>$ of the operator $\hbar A({\hat \eta^+} {\hat
\eta})=\hbar {\hat \eta^+} {\hat \eta}$ and that deviates in general
from this value on superposed states $|F>$ and of course under a
different choice of $\hat{S}=\hbar A({\hat \eta^+} {\hat \eta}) \neq
\hbar {\hat \eta^+} {\hat \eta}$. In order to study the variation of
the action functional on superposed states one need more details on
geometry of their superposition.

In fact only finite, say, $N$ elementary quantum states (EQS's)
($|\hbar 0>, |\hbar 1>,...,|\hbar (N-1)>$) may be involved in the
coherent superposition $|F>$. Then $\cal{H}=C^N$ and the ray space
$CP(\infty)$ will be restricted to finite dimensional $CP(N-1)$.
Hereafter we will use the indices as follows: $0\leq a,b \leq N$,
and $1\leq i,k,m,n,s \leq N-1$. This superposition physically
corresponds to the complete amplitude of some process of the quantum
motion. Sometimes it may be interpreted as a extremal of action
functional of some classical variational problem.

\section{Non-linear treatment of the eigen-problem}
In the famous first report, Schr\"odinger formulated a conditional
variation problem for a ``wave function" corresponding to an
electron in the Coulomb field. This is the eigen-problem for
self-adjoint Hamiltonian in separable Hilbert space \cite{Schr4}.
Starting from the internal symmetry given by $SU(N)$ group we will
discuss variation problem formulated on finite dimension Hilbert
space $\mathcal{H}=C^N$. In this case the eigen-values of some
self-adjoint linear operator $\hat{D}$ are minimal (stationary)
values of the quadratic form $<F|\hat{D}|F>$ on $CP(N-1)$
\cite{Gelfand} achievable at corresponding eigen-vectors of rays.
For us will be important the explicit parametrization of these rays
by local coordinates in $CP(N-1)$ .

Let me start from the eigen-problem for linear operator $\hat{D}$
acting in $C^N$. This is the typically linear problem in the
framework of the theory of linear operators. Furthermore, physicists
have essential simplification of the problem dealing with hermitian
or unitary operators. The invariant sub-spaces of these operators
are one-dimensional even for degenerated eigen-values, the Jordan
cells are one-dimension, and, therefore, these operators are
diagonalizable. If so, there is a reasonable question: why one
should deep into the non-linear formulation, if the linear one seems
to be good enough? It is really so if one deals with the eigen
problem for a single matrix of operator. But I seek for the
dynamical description of the creation-annihilation processes of
particles. This dynamical picture gives a parameterized family of
operators; hence arise the instability problem of the spectrum
structure close to the degeneracy and the Berry's anholonomy too
\cite{Berry198}. Therefore the non-linear approach is in fact
inevitable since the realistic theory in any case requires the
dynamical re-definitions of the creation-annihilation operators (the
most clear this is realized by the $Y_{ni}, Z_{ni}$ coefficients of
Dirac \cite{Dirac1}).

The quantum mechanics assumes the priority of the Hamiltonian given
by some classical model which henceforth should be ``quantized". It
is known that this procedure is ambiquous. In order to avoid the
ambiguity, I intend to use a {\it quantum state} itself and the
invariant conditions of its conservation and perturbation. These
invariant conditions are rooted into the global geometry of the
dynamical group manifold. Namely, the geometry of $G=SU(N)$, the
isotropy group $H=U(1)\times U(N-1)$ of the pure quantum state, and
the coset $G/H=SU(N)/U(1)\times U(N-1)$ geometry, play an essential
role in the quantum state evolution (the super-relativity principle
\cite{Le2}). The stationary states (some eigen-states of action
operator, i.e. the states of motion with the least action) may be
treated as {\it initial conditions} for GCS evolution. Particulary
they may represent a local minimum of energy (vacuum).

Let me assume that $\left\{|\hbar a>\right\}_0^{N-1}$ is the basis
in Hilbert space $\mathcal{H}$. Then a typical vector $|F> \in
\mathcal{H}$ may be represented as a superposition
$|F>=\sum_0^{N-1}f^a|\hbar a>$. The eigen-problem may be formulated
for some hermitian dynamical variable $\hat{D}$ on these typical
vectors $\hat{D}|F>=\lambda_D|F> $. This equation may be written in
components as follows: $\sum_0^{N-1}D^a_b f^b=\lambda_D f^a$, where
$D^a_b=<a|\hat{D}|b>$.
\begin{equation}
\hat{D}=\sum_{a,b \geq 0} <a|\hat{D}|b> \hat{P}_{ab}=\sum_{a,b \geq
0}D_{ab}\hat{P}_{ab}=\sum_{a,b \geq 0} F^{\alpha}_D
\hat{\lambda}_{\alpha,(ab)} \hat{P}_{ab},
\end{equation}
where projector is as follows
\begin{equation}
\hat{P}_{ab}=|a><b|=\frac{1}{\sqrt{a! b!}}:(\eta^+)^a \exp(-\eta^+
\eta)(\eta)^b:,
\end{equation}
and where symbol $:...:$ means the the normal ordering of operators,
and functions $F^{\alpha}_D$ obey some field equations which will be
discussed later. In particular, the Hamiltonian has similar
representation
\begin{equation}
\hat{H}=\sum_{a,b \geq 0} <a|\hat{H}|b> \hat{P}_{ab}=\sum_{a,b \geq
0}H_{ab}\hat{P}_{ab}=\hbar \sum_{a,b \geq 0} \Omega^{\alpha}
\hat{\lambda}_{\alpha,(ab)} \hat{P}_{ab},
\end{equation}
i.e $F^{\alpha}_H=\hbar \Omega^{\alpha}$ \cite{Le3}.

One has the spectrum of $\lambda_D:
\left\{\lambda_0,...,\lambda_{N-1} \right\}$ from the equation
$Det(\hat{D}-\lambda_D \hat{E})=0$ , and then one has the set of
equations $\hat{D}|D_p>=\lambda_p|D_p> $, where $p=0,...,N-1$ and
$|D_p>=\sum_0^{N-1}g^a_p|\hbar a>$ are eigen-vectors. It is worse
while to note here that the solution of this problem gives rather
rays than vectors, since eigen-vectors are defined up to the complex
factor. In other words we deal with rays or points of the non-linear
complex projective space $CP(N-1)$ for $N \times N$ matrix of the
linear operator acting on $C^N$. The Hilbert spaces of the infinite
dimension will be discussed later.

For each eigen-vector $ |D_p> $ corresponding $\lambda_p$ it is
possible to chose at least one such component $g^j_p$ of the
$|D_p>$, that $|g^j_p| \neq 0$. This choice defines in fact the map
$U_{j(p)}$ of the local projective coordinates for each
eigen-vecrtor
\begin{equation}
\pi^i_{j(p)}=\cases{\frac{g^i_p}{g^j_p},&if $ 1 \leq i < j$ \cr
\frac{g^{i+1}_p}{g^j_p}&if $j \leq i < N-1$}
\end{equation}\label{coor}
of the ray corresponding $|D_p>$ in $CP(N-1)$. Note, if all
$\pi^i_{j(p)}=0$ it means that one has the ``pure" state
$|D_p>=g^j_p|j>$ (without summation in $j$). Any different points of
the $CP(N-1)$ corresponds to the GCS's. They will be treated as
self-rays of some deformed action operator. Beside this I will treat
the superposition state $|G>=\sum_{a=0}^{N-1} g^a|a\hbar>$ as
``analytic continuation" of the of eigen-vector for an arbitrary set
of the local coordinates. Our aim is to find equations for these
operators coinciding with initial operator at original self-ray
(initial conditions) given by GCS.

People frequently omit the index $p$, assuming that
$\lambda:=\lambda_p$, for $j=0$. Then they have, say, for the
Hamiltonian matrix
\begin{eqnarray}
\hat{H} = \left( \matrix {H_{00} & H_{01} &... H_{0i} &... H_{0N-1}
\cr H_{10} & H_{11} &... H_{1 i} &... H_{1 N-1}  \cr H_{20} & H_{21}
&... H_{2 i} &... H_{2 N-1}  \cr . \cr . \cr . \cr H_{N-1 0} &
H_{N-1 1} &... H_{N-1 i} &... H_{N-1 N-1} } \right)
\end{eqnarray}
the eigen-problem
\begin{eqnarray}
(H_{00}-\lambda)\psi^0 + H_{01}\psi^1 +...+ H_{0i}\psi^i +...+
H_{0N-1}\psi^{N-1} =0 \cr H_{10}\psi^0 + (H_{11}-\lambda )\psi^1
+...+ H_{1 i}\psi^i +...+ H_{1 N-1}\psi^{N-1} =0  \cr H_{20}\psi^0 +
H_{21}\psi^1 + (H_{22}-\lambda)\psi^2 +...+ H_{2 i}\psi^i +...+ H_{2
N-1}\psi^{N-1} =0 \cr . \cr . \cr . \cr H_{N-1 0}\psi^0 + H_{N-1
1}\psi^1 +...+ H_{N-1 i}\psi^i +...+ (H_{N-1 N-1} - \lambda)
\psi^{N-1} =0,
\end{eqnarray}
where $\psi^a:=g^a_0$.

In accordance with our assumption the $\lambda$ is such that $\psi^0
\neq 0$. Let then divide all equations by $\psi^0$. Introducing
local coordinates $\pi^i=\frac{\psi^i}{\psi^0}$, we get the system
of the non-homogeneous equations
\begin{eqnarray}
(H_{11}-\lambda )\pi^1 +...+ H_{1 i}\pi^i +...+ H_{1 N-1}\pi^{N-1}
=- H_{10}  \cr H_{21}\pi^1 + (H_{22}-\lambda)\pi^2 +...+ H_{2
i}\pi^i +...+ H_{2 N-1}\pi^{N-1} =- H_{20} \cr . \cr . \cr . \cr
H_{N-1 1}\pi^1 +...+ H_{N-1 i}\pi^i +...+ (H_{N-1 N-1} - \lambda)
\pi^{N-1} = -H_{N-1 0},
\end{eqnarray}
where the first equation
\begin{eqnarray}
H_{01}\pi^1 +...+ H_{0i}\pi^i +...+ H_{0 N-1}\pi^{N-1}
=-(H_{00}-\lambda)
\end{eqnarray}
is omitted. If $D=det(H_{ik}-\lambda \delta_{ik}) \neq 0, i \neq 0,
k \neq 0$ then the single defined solutions of this system may be
expressed through the Cramer's rule
\begin{eqnarray}
\pi^1=\frac{D_1}{D},...,\pi^{N-1}=\frac{D_{N-1}}{D}.
\end{eqnarray}
It is easy to see that these solutions being substituted into the
first omitted equation give us simply re-formulated initial
characteristic equation of the eigen-problem. Therefore one has the
single value nonlinear solution of the eigen-problem instead of the
linear one with additional freedom of a complex scale
multiplication.

This approach does not give essential advantage for a single
operator and it only shows that the formulation in local coordinates
is quite natural. But if one tries to understand how the
multi-dimensional variation of the hermitian operator included in a
parameterized family, the local formulation is in inevitable. First
of all it is interesting to know the invariants of such variations.
In particular the quantum measurement of dynamical variable
represented by hermitian $N \times N$ matrix should be described in
the spirit of typical polarization measurement of the coherent
photons \cite{Le1}. I will put below the sketch that depicts the
operational ``travel" of polarization state on the Poincar\'e
sphere.

The initial state $|x>$ is modulated passing through an optically
active medium (say using the Faraday effect in YIG film magnetized
along the main axes in the $z$-direction by a harmonic magnetic
field with frequency $\Omega$ and the angle amplitude $\beta$).
Formally this process may be described by the action of the unitary
matrix ${\hat h}_{os_3}$ belonging to the isotropy group of $|R>$
\cite{Le2}. Then the coherence vector will oscillate along the
equator of the Poincar\'e sphere. The next step is the dragging of
the oscillating state $|x'(t)>=\hat{h_{os_3}}|x>$ with frequency
$\omega$ up to the ``north pole'' corresponding to the state $|R>$.
In fact this is the motion of the coherence vector. This may be
achieved by the variation of the azimuth of the linear polarized
state from $\frac{\theta}{2}=-\frac{\pi}{4}$ up to
$\frac{\theta}{2}=\frac{\pi}{4}$ with help of the dense flint of
appropriate length embedded into the sweeping magnetic field.
Further this beam should pass the $\lambda /4$ plate. This process
of variation of the ellipticity of the polarization ellipse may be
described by the unitary matrix ${\hat b}_{os'_1}$ belonging to the
coset homogeneous sub-manifold $U(2)/U(1) \times U(1)=CP(1)$ of the
dynamical group $U(2)$ \cite{Le2}. This dragging without modulation
leads to the evolution of the initial state along the geodesic of
$CP(1)$ and the trace of the coherent vector is the meridian of the
Poincar\'e sphere between the equator and one of the poles. The
modulation deforms both the geodesic and the corresponding trace of
the coherence vector on the Poincar\'e sphere during such unitary
evolution.

The action of the $\lambda /4$ plate depends upon the state of the
incoming beam (the relative orientation of the fast axes of the
plate and the polarization of the beam). Furthermore, only relative
phases and amplitudes of photons in the beam have a physical meaning
for the $\lambda /4$ plate. Neither the absolute amplitude
(intensity of the beam), nor the general phase affect the
polarization character of the outgoing state. It means that the
device action depends only upon the local coordinates $\pi^1
=\frac{\Psi^1}{\Psi^0} \in CP(1)$. Small relative re-orientation of
the $\lambda /4$ plate and the incoming beam leads to a small
variation of the outgoing state. This means that the $\lambda /4$
plate re-orientation generates the tangent vector to $CP(1)$. It is
natural to discuss the two components of such a vector: velocities
of the variations of the ellipticity and of the azimuth
(inclination) angle of the polarization ellipse. They are  examples
of LDV. The comparison of such dynamical variables for different
coherent states requires that affine parallel transport agrees with
the Fubini-Study metric.

As far as I know the generalized problem of the quantum measurement
of an arbitrary hermitian dynamical variable $\hat{H}=E^{\alpha}
\hat{\lambda}_{\alpha}, \quad \hat{\lambda}_{\alpha} \in AlgSU(N)$
in the operational manner given above was newer done. It is solved
here by the exact analytical diagonalization of an hermitian matrix.
Previously this problem was solved partly in the works
\cite{Onf,Ost,Le4}. Geometrically it looks like embedding ``the
ellipsoid of polarizations" into the iso-space of the adjoint
representation of $SU(N)$. This ellipsoid is associated with the
quadric form $<F|\hat{H}|F>=\sum_1^{N^2-1} E^{\alpha}<F|
\hat{\lambda}_{\alpha}|F>=H_{ab}(E^{\alpha})f^{a*} f^b$ depending on
$N^2-1$ real parameters $E^{\alpha}$. The shape of this ellipsoid
with $N$ main axes is giving by the $2(N-1)$ parameters of the coset
transformations $G/H=SU(N)/S[U(1) \times U(N-1)]=CP(N-1)$ relate to
the $(N-1)$ complex local coordinates of the eigen-state of
$\hat{H}$ in $CP(N-1)$. Its orientation in iso-space $R^{N^2-1}$ is
much more complicated than it was in the case of $R^3$. It is given
by generators of the isotropy group containing $N-1=rank(AlgSU(N))$
independent parameters of ``rotations'' about commutative operators
$\hat{\lambda}_3, \hat{\lambda}_8, \hat{\lambda}_{15},...$ and
$(N-1)(N-2)$ parameters of rotations about non-commutative
operators.  All these $(N-1)^2=(N-1)+(N-1)(N-2)$ gauge angles of the
isotropy group $H=S[U(1) \times U(N-1)]$ of the eigen-state giving
orientation of this ellipsoid in iso-space $R^{N^2-1}$ will be
calculated now during the process of analytical diagonalization of
the hermitian matrix $H_{ab}=<a|\hat{H}|b>$ corresponding to some
dynamical variable $\hat{H}$.

\pagebreak

{\bf Stage 1. Reduction of the general Hermitian Matrix to
three-diagonal form.}

Let me start from general hermitian $N \times N$ matrix
\begin{eqnarray}
\hat{H} = \left( \matrix {H_{00} & H_{01} &... H_{0i} &... H_{0N-1}
\cr H_{10} & H_{11} &... H_{1 i} &... H_{1 N-1}  \cr H_{20} & H_{21}
&... H_{2 i} &... H_{2 N-1}  \cr . \cr . \cr . \cr H_{N-1 0} &
H_{N-1 1} &... H_{N-1 i} &... H_{N-1 N-1} } \right).
\end{eqnarray}
One should choose some basis in $C^N$. I will take the standard
basis
\begin{eqnarray}
|1>= \left( \matrix {1 \cr 0 \cr 0 \cr . \cr . \cr . \cr 0 }
\right),&|2>= \left( \matrix {0 \cr 1 \cr 0 \cr . \cr . \cr . \cr 0
} \right),...,|N>= \left( \matrix {0 \cr 0 \cr 0 \cr . \cr . \cr .
\cr 1 } \right).
\end{eqnarray}
Now if one choose, say, $N=3$ then the standard Gell-Mann
$\hat{\lambda}$ matrices may be distinguished into the two sets in
respect with for example the state $|1>$ : B-set $\hat{\lambda_1},
\hat{\lambda_2}, \hat{\lambda_4}, \hat{\lambda_5}$ whose exponents
act effectively on the $|1>$, and the H-set $\hat{\lambda_3},
\hat{\lambda_8}, \hat{\lambda_6}, \hat{\lambda_7}$ , whose exponents
that leave $|1>$ intact. For any finite dimension $N$ one may define
the ``I-spin" $(1 \leq I \leq N)$ as an analog of the well known
``T-,U-,V- spins" of the $SU(3)$ theory using the invariant
character of the commutation relations of B-and H-sets
\begin{eqnarray}
[B,B] \in H, \quad [H,H] \in H, \quad [B,H] \in B.
\end{eqnarray}
Let me now to represent our hermitian matrix in following form
\begin{eqnarray}
\hat{H} = \left( \matrix {0 & H_{01} &... H_{0i} &... H_{0N-1}  \cr
H_{10} & 0 &... 0 &... 0  \cr H_{20} & 0 &... 0 &... 0 \cr . \cr .
\cr . \cr H_{N-1 0} & 0 &... 0 &... 0 } \right)_B \cr + \left(
\matrix {H_{00} & 0 &... 0 &... 0  \cr 0 & H_{11} &... H_{1 i} &...
H_{1 N-1} \cr 0 & H_{21} &... H_{2 i} &... H_{2 N-1}  \cr . \cr .
\cr . \cr 0 & H_{N-1 1} &... H_{N-1 i} &... H_{N-1 N-1} } \right)_H.
\end{eqnarray}
In respect with ket $|1>$ one may to classify the first matrix as
$B-type$ and the second one as a matrix of the $H-type$. I will
apply now the ``squeezing ansatz" \cite{Le2,Le4}.  The first
``squeezing'' unitary matrix is
\begin{equation}
\hat{U}_1= \left( \matrix{1&0&0&.&.&.&0 \cr 0&1&0&.&.&.&0 \cr
.&.&.&.&.&.&. \cr .&.&.&.&.&.&. \cr 0&.&.&.&1&0&0 \cr .&.&.&.&0&\cos
\phi_1&e^{i\psi_1} \sin \phi_1 \cr 0&0&.&.&0&-e^{-i\psi_1} \sin
\phi_1&\cos \phi_1 } \right ).
\end{equation}
The transformation of similarity being applied to our matrix gives
$\hat{H}_1 =\hat{U}_1^+ \hat{H} \hat{U}_1$ with the result for
$\hat{H}_B$ shown for simplicity in the case $N=4$
\begin{eqnarray}
\hat{H}_{B1} = \left( \matrix {0&H_{01}&\tilde{H_{02}}
&\tilde{H_{03}} \cr H_{01}^*&0&0&0\cr \tilde{H_{02}^*} &0&0&0 \cr
\tilde{H_{03}^*} &0&0&0} \right),
\end{eqnarray}
where $\tilde{H_{02}}=H_{02} \cos \phi-H_{03}\sin \phi e^{-i\psi}$
and $\tilde{H_{03}} =H_{02}\sin \phi e^{i\psi} + H_{03} \cos \phi $.
Now one has solve two ``equations of annihilation'' of $\Re (H_{02}
\sin(\phi)e^{i\psi} + H_{03}\cos(\phi))=0$ and $\Im (H_{02}
\sin(\phi)e^{i\psi} + H_{03}\cos(\phi))=0$ in order to eliminate the
last element of the first row and its hermitian conjugate
\cite{Le4}. This gives us $\phi'_1$ and $\psi'_1$. I will put
$H_{02}=\alpha_{02}+i\beta_{02}$ and
$H_{03}=\alpha_{03}+i\beta_{03}$, then the solution of the
``equations of annihilation'' is as follows:
\begin{eqnarray}
\phi'_1=\arctan
\sqrt{\frac{\alpha_{03}^2+\beta_{03}^2}{\alpha_{02}^2+\beta_{02}^2}},
\cr \psi'_1=\arctan \frac{\alpha_{03} \beta_{02} - \alpha_{02}
\beta_{03}}
{\sqrt{(\alpha_{02}^2+\beta_{02}^2)(\alpha_{03}^2+\beta_{03}^2)}}.
\end{eqnarray}
This transformation acts of course on the second matrix $\hat{H}_H$,
but it easy to see that its structure is intact. The next step is
the similarity transformations given by the matrix with the
diagonally shifted transformation block
\begin{equation}
\hat{U}_2= \left( \matrix{ 1&0&0&.&.&.&0 \cr 0&1&0&.&.&.&0 \cr
.&.&.&.&.&.&. \cr 0&.&.&.&1&0&0 \cr .&.&.&.&0&cos \phi_2&e^{i\psi_2}
sin \phi_2 \cr 0&0&.&.&0&-e^{-i\psi_2}sin \phi_2&cos \phi_2 \cr
0&.&.&.&0&0&1 } \right )
\end{equation}
and the similar evaluation of $\psi'_2$, $\phi'_2$. Generally one
should make $N-2$ steps in order to annulate $N-2$ elements of the
first row. The next step is to represent our transformed $\hat{H}_1
=\hat{U}_1^+ \hat{H} \hat{U}_1$ as follows:
\begin{eqnarray}
\hat{H}_1 = \left( \matrix {0 & \tilde{H}_{01} &0 &... 0 \cr
\tilde{H}_{10} & 0 &\tilde{H}_{12}&... \tilde{H}_{1 N-1} \cr 0 &
\tilde{H}_{21} &0 &... 0 \cr . \cr . \cr . \cr 0 & \tilde{H}_{N-1,1}
&... 0 &... 0 } \right)_B \cr + \left( \matrix {H_{00} & 0 &... 0
&... 0 \cr 0 & \tilde{H}_{11} &...0 &... 0\cr 0 & 0 &...
\tilde{H}_{2 i} &... \tilde{H}_{2 N-1} \cr . \cr . \cr . \cr 0 & 0
&... \tilde{H}_{N-1,i} &... \tilde{H}_{N-1,N-1} } \right)_H.
\end{eqnarray}
Now one should applied the squeezing ansatz in $N-3$ steps for
second row, etc., generally one has $(N-1)(N-2)$ orientation angles.
Thereby we come to the three-diagonal form of the our matrix.

{\bf Stage 2. Diagonalization of the three-diagonal form.}

The eigen-problem for the three-diagonal hermitian matrix is well
know, but I will put it here for the completeness. The eigen-
problem $(\hat{\tilde{H}} - \lambda \hat{E})|\xi>=0$ for the
three-diagonal matrix has the following form
\begin{eqnarray}
\left( \matrix {\tilde{H}_{00}\xi^0&\tilde{H}_{01}\xi^1&0&.&.&.&0
\cr
\tilde{H}_{01}^*\xi^0&\tilde{H}_{11}\xi^1&\tilde{H}_{12}\xi^2&0&.&.&0\cr
0&\tilde{H}_{12}^*\xi^1&\tilde{H}_{22}\xi^2&\tilde{H}_{23}\xi^3&.&.&0
\cr 0&0& \tilde{H}_{23}^*\xi^2&.&.&.&0 \cr 0&0&.&.&.&.&0 \cr
0&0&.&.&.&.&\tilde{H}_{N-1,N-2}\xi^{N-1} \cr
0&0&.&.&.&.&\tilde{H}_{N-1,N-1}\xi^{N-1}} \right) = \left( \matrix
{\lambda \xi^0 \cr \lambda \xi^1 \cr \lambda \xi^2 \cr . \cr . \cr .
\cr \lambda \xi^{N-1}} \right).
\end{eqnarray}
Since $\xi^1=\frac{\lambda -\tilde{H}_{00}}{\tilde{H}_{01}} \xi^0$,
etc., one has the reccurrent relations between all components of the
eigen-vector corresponding to given $\lambda$. Thereby only $N-1$
complex local coordinates
$(\pi^1=\frac{\xi^1}{\xi^0},...,\pi^{N-1}=\frac{\xi^{N-1}}{\xi^0})$
giving the shape of the ellipsoid of polarization have invariant
sense as it was mentioned above.

{\bf Stage 3. The coset ``force'' acting during a measurement}

The real measurement assumes some interaction of the measurement
device and incoming state. If we assume for simplicity that incoming
state is $|1>$ (modulation, etc. are neglected), then all
transformations from $H$-subalgebra will leave it intact. Only the
coset unitary transformations \pagebreak
\begin{eqnarray}
& \hat{T}(t,g) = \cr & \left( \matrix{\cos gt&\frac{-p^{1*}}{g} \sin
gt&\frac{-p^{2*}}{g}\sin gt&.&\frac{-p^{N-1*}}{g}\sin gt \cr
\frac{p^1}{g} \sin gt&1+[\frac{|p^1|}{g}]^2 (\cos gt -1)&[\frac{p^1
p^{2*}}{g}]^2 (\cos gt -1)&.&[\frac{p^1 p^{N-1*}}{g}]^2 (\cos gt -1)
\cr .&.&.&.&. \cr .&.&.&.&. \cr .&.&.&.&. \cr \frac{p^{N-1}}{g}\sin
gt&[\frac{p^{1*} p^{N-1}}{g}]^2 (\cos gt
-1)&.&.&1+[\frac{|p^{N-1}|}{g}]^2 (\cos gt -1)} \right),
\end{eqnarray}
where $g=\sqrt{|p^1|^2+,...,+|p^{N-1}|^2}$ will effectively to
variate this state dragging it along one of the geodesic in
$CP(N-1)$ \cite{Le2}. This matrix describe the process of the
transition from one pure state to another, in particular between two
eigen-states connected by the geodesic. This means that these
transformations deform the ellipsoid. All possible shapes of these
ellipsoids are distributed along a single geodesic.

Generally, in the dynamical situation this ``stationary'' global
procedure is not applicable and one should go to the local analog of
$\hat{\lambda}$-matrices, i.e. $SU(N)$ generators and related
dynamical variables should be parameterized by the local quantum
states coordinates $(\pi^1,...,\pi^{N-1})$.


\section{Local dynamical variables} The state space ${\cal H}$
with finite action quanta is a stationary construction. We introduce
dynamics {\it by the velocities of the GCS variation} representing
some ``elementary excitations'' (quantum particles). Their dynamics
is specified by the Hamiltonian, giving time variation velocities of
the action quantum numbers in different directions of the tangent
Hilbert space $T_{(\pi^1,...,\pi^{N-1})} CP(N-1)$ which takes the
place of the ordinary linear quantum state space as will be
explained below. The rate of the action variation gives the energy
of the ``particles'' whose expression should be established by some
wave equations.

The local dynamical variables correspond to the internal $SU(N)$
group of the GCS and their breakdown should be expressed now in
terms of the local coordinates $\pi^k$. The Fubini-Study metric
\begin{equation}
G_{ik^*} = [(1+ \sum |\pi^s|^2) \delta_{ik}- \pi^{i^*} \pi^k](1+
\sum |\pi^s|^2)^{-2} \label{FS}
\end{equation}
and the affine connection
\begin{eqnarray}
\Gamma^i_{mn} = \frac{1}{2}G^{ip^*} (\frac{\partial
G_{mp^*}}{\partial \pi^n} + \frac{\partial G_{p^*n}}{\partial
\pi^m}) = -  \frac{\delta^i_m \pi^{n^*} + \delta^i_n \pi^{m^*}}{1+
\sum |\pi^s|^2} \label{Gamma}
\end{eqnarray}
in these coordinates will be used. Hence the internal dynamical
variables and their norms should be state-dependent, i.e. local in
the state space \cite{Le1,Le2}. These local dynamical variables
realize a non-linear representation of the unitary global $SU(N)$
group in the Hilbert state space $C^N$. Namely, $N^2-1$ generators
of $G = SU(N)$ may be divided in accordance with the Cartan
decomposition: $[B,B] \in H, [B,H] \in B, [B,B] \in H$. The
$(N-1)^2$ generators
\begin{eqnarray}
\Phi_h^i \frac{\partial}{\partial \pi^i}+c.c. \in H,\quad 1 \le h
\le (N-1)^2
\end{eqnarray}
of the isotropy group $H = U(1)\times U(N-1)$ of the ray (Cartan
sub-algebra) and $2(N-1)$ generators
\begin{eqnarray}
\Phi_b^i \frac{\partial}{\partial \pi^i} + c.c. \in B, \quad 1 \le b
\le 2(N-1)
\end{eqnarray}
are the coset $G/H = SU(N)/S[U(1) \times U(N-1)]$ generators
realizing the breakdown of the $G = SU(N)$ symmetry of the GCS.
Furthermore, the $(N-1)^2$ generators of the Cartan sub-algebra may
be divided into the two sets of operators: $1 \le c \le N-1$ ($N-1$
is the rank of $Alg SU(N)$) Abelian operators, and $1 \le q \le
(N-1)(N-2)$ non-Abelian operators corresponding to the
non-commutative part of the Cartan sub-algebra of the isotropy
(gauge) group. Here $\Phi^i_{\sigma}, \quad 1 \le \sigma \le N^2-1 $
are the coefficient functions of the generators of the non-linear
$SU(N)$ realization. They give the infinitesimal shift of the
$i$-component of the coherent state driven by the $\sigma$-component
of the unitary multipole field $\Omega^{\alpha}$ rotating the
generators of $Alg SU(N)$ and they are defined as follows:
\begin{equation}
\Phi_{\sigma}^i = \lim_{\epsilon \to 0} \epsilon^{-1}
\biggl\{\frac{[\exp(i\epsilon \lambda_{\sigma})]_m^i g^m}{[\exp(i
\epsilon \lambda_{\sigma})]_m^j g^m }-\frac{g^i}{g^j} \biggr\}=
\lim_{\epsilon \to 0} \epsilon^{-1} \{ \pi^i(\epsilon
\lambda_{\sigma}) -\pi^i \},
\end{equation}
\cite{Le2}. Then the sum of the $N^2-1$ the energies associated with
intensity of deformations of the GCS is represented  by the local
Hamiltonian vector field $\vec{H}$ which is linear in the partial
derivatives $\frac{\partial }{\partial \pi^i} = \frac{1}{2}
(\frac{\partial }{\partial \Re{\pi^i}} - i \frac{\partial }{\partial
\Im{\pi^i}})$ and $\frac{\partial }{\partial \pi^{*i}} = \frac{1}{2}
(\frac{\partial }{\partial \Re{\pi^i}} + i \frac{\partial }{\partial
\Im{\pi^i}})$. In other words it is the tangent vector to $CP(N-1)$
\begin{eqnarray}
\vec{H}=T_c+T_q +V_b = \hbar \Omega^c \Phi_c^i \frac{\partial
}{\partial \pi^i} + \hbar \Omega^q \Phi_q^i \frac{\partial
}{\partial \pi^i} + \hbar \Omega^b \Phi_b^i \frac{\partial
}{\partial \pi^i} + c.c. \label{field}
\end{eqnarray}
The characteristic equations for the PDE $\vec{H}|E>=E|E>$ give the
parametric representations of their solutions in $CP(N-1)$. We will
identify the parameter $\tau$ in these equations with a ``universal
time of evolution'' such as the world time \cite{H1}. This time is
the measure of the GCS variation, i.e. it is a measure of the
distance in $CP(N-1)$ (the length of the evolution trajectory in the
Fubini-Study metric) expressed in time units. The energy
quantization will be discussed elsewhere.

In order to express some eigen-vector in the local coordinates, I
put
\begin{eqnarray}
|D_p(\pi^1_{j(p)},...,\pi^{N-1}_{j(p)})>
 =\sum_0^{N-1}
g^a(\pi^1_{j(p)},...,\pi^{N-1}_{j(p)})|a>,
\end{eqnarray}
where $\sum_{a=0}^{N-1} |g^a|^2= R^2$, and
\begin{eqnarray}
g^0(\pi^1_{j(p)},...,\pi^{N-1}_{j(p)})=\frac{R^2}{\sqrt{R^2+
\sum_{s=1}^{N-1}|\pi^s_{j(p)}|^2}}.
\end{eqnarray}
For $1\leq i\leq N-1$ one has
\begin{eqnarray}
g^i(\pi^1_{j(p)},...,\pi^{N-1}_{j(p)})=\frac{R
\pi^i_{j(p)}}{\sqrt{R^2+\sum_{s=1}^{N-1}|\pi^s_{j(p)}|^2}},
\end{eqnarray}
i.e. $CP(N-1)$ is embedded in the Hilbert space ${\cal{H}}=C^N$.
Hereafter I will suppose $R=1$.

Now we see that all eigen-vectors corresponding to different
eigen-values (even under the degeneration) are applied to different
points $(\pi^1_{j(p)},...,\pi^{N-1}_{j(p)})$ of the $CP(N-1)$.
Nevertheless the eigen-vectors
$|D_p(\pi^1_{j(p)},...,\pi^{N-1}_{j(p)})>$ are mutually orthogonal
in ${\cal{H}}=C^N$ if $\hat{H}$ is hermitian Hamiltonian. Therefore
one has the ``splitting" or delocalization of degenerated
eigen-states in $CP(N-1)$. Thus the local coordinates $\pi^i$  gives
the convenient parametrization of the $SU(N)$ action as one will see
below.

Let me assume that $|G>=\sum_{a=0}^{N-1} g^a|a\hbar>$ is a ``ground
state" of some the least action problem. Then the velocity of the
ground state evolution relative world time is given by the formula
\begin{eqnarray}\label{41}
|H> = \frac{d|G>}{d\tau}=\frac{\partial g^a}{\partial
\pi^i}\frac{d\pi^i}{d\tau}|a\hbar>=|T_i>\frac{d\pi^i}{d\tau}=H^i|T_i>,
\end{eqnarray}
 is the tangent vector to the evolution curve
$\pi^i=\pi^i(\tau)$, where
\begin{eqnarray}\label{42}
|T_i> = \frac{\partial g^a}{\partial \pi^i}|a\hbar>=T^a_i|a\hbar>.
\end{eqnarray}
Then the ``acceleration'' is as follows
\begin{eqnarray}\label{43}
|A> =
\frac{d^2|G>}{d\tau^2}=|g_{ik}>\frac{d\pi^i}{d\tau}\frac{d\pi^k}{d\tau}
+|T_i>\frac{d^2\pi^i}{d\tau^2}=|N_{ik}>\frac{d\pi^i}{d\tau}\frac{d\pi^k}{d\tau}\cr
+(\frac{d^2\pi^s}{d\tau^2}+\Gamma_{ik}^s
\frac{d\pi^i}{d\tau}\frac{d\pi^k}{d\tau})|T_s>,
\end{eqnarray}
where
\begin{eqnarray}\label{44}
|g_{ik}>=\frac{\partial^2 g^a}{\partial \pi^i \partial \pi^k}
|a\hbar>=|N_{ik}>+\Gamma_{ik}^s|T_s>
\end{eqnarray}
and the state
\begin{eqnarray}\label{45}
|N> = N^a|a\hbar>=(\frac{\partial^2 g^a}{\partial \pi^i \partial
\pi^k}-\Gamma_{ik}^s \frac{\partial g^a}{\partial \pi^s})
\frac{d\pi^i}{d\tau}\frac{d\pi^k}{d\tau}|a\hbar>
\end{eqnarray}
is the normal to the ``hypersurface'' of the ground states. Then the
minimization of this ``acceleration'' under the transition from
point $\tau$ to $\tau+d\tau$ may be achieved by the annihilation of
the tangential component
\begin{equation}
(\frac{d^2\pi^s}{d\tau^2}+\Gamma_{ik}^s
\frac{d\pi^i}{d\tau}\frac{d\pi^k}{d\tau})|T_s>=0
\end{equation}
i.e. under the condition of the affine parallel transport of the
Hamiltonian vector field
\begin{equation}\label{par_tr}
dH^s +\Gamma^s_{ik}H^id\pi^k =0.
\end{equation}

The derivatives arose in previous formulas are as following:
\begin{eqnarray}
T^0_i=\frac{\partial g^0}{\partial \pi^i}=-\frac{1}{2} \frac{
 \pi^{*i}}{\left(\sqrt{\sum_{s=1}^{N-1} |\pi^s|^2+1}\right)^3},\cr
T^{0*}_k=\frac{\partial g^0}{\partial \pi^{*k}}=-\frac{1}{2} \frac{
\pi^{k}}{\left(\sqrt{\sum_{s=1}^{N-1} |\pi^s|^2+1}\right)^3}
\end{eqnarray}
and for other components ($a \geq 1$) one has
\begin{eqnarray}
T^m_i=\frac{\partial g^m}{\partial \pi^i}     & = &
\left(\frac{\delta^m_i}{\sqrt{\sum_{s=1}^{N-1} |\pi^s|^2+1}}-
\frac{1}{2} \frac{\pi^m \pi^{*i}}
{\left(\sqrt{\sum_{s=1}^{N-1} |\pi^s|^2+1}\right)^3}\right), \nonumber \\
T^{m*}_k=\frac{\partial g^{*m}}{\partial \pi^{*k}} & = &
\left(\frac{\delta^m_k}{\sqrt{\sum_{s=1}^{N-1} |\pi^s|^2+1}}-
\frac{1}{2} \frac{\pi^{*m} \pi^{k}} {\left(\sqrt{\sum_{s=1}^{N-1}
|\pi^s|^2+1}\right)^3}\right).
\end{eqnarray}

At first sight they have neither connection with space-time (or
momentum) representation which we need for a dynamical description.
This representation one should obtain by the introduction of a local
{\it dynamical space-time} \cite{Le5,Le6}.

We saw that $SU(N)$ geometry gives the shape and the orientation of
the ellipsoid associated with the ``average" of dynamical variable
given by a quadric form $<F|\hat{D}|F>$. If it is taking ``as given"
it show only primitive eigen-value problem. But if one rises the
question about real operational sense of the quantum measurement of
this dynamical variable or the process of the transition from one
eigen-state to another, one sees that quantum state and dynamical
variable involved in much more complicated relations that it is
given in the orthodox quantum scheme. The simple reason for this is
that the decomposition (representation) of the state vector of a
quantum system strongly depends on the spectrum and eigen-vectors of
its dynamical variable. Overloaded system of the GCS's supplies us
by enough big ``reserve'' of functions but their superposition
should be local and they span a tangent space at any specific point
of $CP(N-1)$ marked by the local coordinates.

The ``probability" may be introduced now by pure geometric way like
$cos^2 \phi $ in tangent state space as follows.

For any two tangent vectors $D_1^i=<D_1|T_i>, D_2^i=<D_2|T_i>$ one
can define the scalar product
\begin{eqnarray}\label{}
(D_1,D_2)=\Re G_{ik^*} D_1^i D_2^{k^*}=\cos \phi_{1,2}
(D_1,D_1)^{1/2} (D_2,D_2)^{1/2}.
\end{eqnarray}
Then the value
\begin{eqnarray}\label{}
P_{1,2}(\pi^1_{j(p)},...,\pi^{N-1}_{j(p)})=\cos^2
\phi_{1,2}=\frac{(D_1,D_2)^2}{(D_1,D_1) (D_2,D_2)}
\end{eqnarray}
may be treated as a relative probability of the appearance of two
states arising during the measurements of two different dynamical
variables $D_1, D_2$ by the variation of the initial GCS
$(\pi^1_{j(p)},...,\pi^{N-1}_{j(p)})$.

Some LDV $\vec{\Psi}=\Psi^i \frac{\partial}{\partial \pi^i} + c.c. $
may be associated with the ``state vector"  $|\Psi> \in \mathcal{H}$
which has tangent components $\Psi^i=<T_i|\Psi>$ in
$T_{\pi}CP(N-1)$. Thus the scalar product
\begin{eqnarray}\label{}
(\Psi,D)=\Re G_{ik^*} \Psi^i D^{k^*}=\cos \phi_{\Psi,D}
(\Psi,\Psi)^{1/2} (D,D)^{1/2}.
\end{eqnarray}
gives the local correlation between two LDV's at same GCS. The
cosines of directions
\begin{eqnarray}\label{}
P_{\Psi,i}(\pi^1_{j(p)},...,\pi^{N-1}_{j(p)})=\cos^2
\phi_{\Psi,i}=\frac{(\Psi,D^i)^2}{(\Psi,\Psi) (D^i,D^i)}
\end{eqnarray}
may be identified with ``probabilities" in each tangent direction of
$T_{\pi}CP(N-1)$. The conservation law of ``probability" is given by
the simple identity
\begin{eqnarray}\label{}
\sum_{i=1}^{N-1} P_{\Psi,i}=\sum_{i=1}^{N-1}\cos^2 \phi_{\Psi,i}= 1.
\end{eqnarray}

The notion of the ``probability" is of course justified by our
experience since different kinds of fluctuations prevent the exact
knowledge of any quantum dynamical variable. That is not only
because the uncertainty relation between {\it two} canonically
conjugated dynamical variables puts the limit of accuracy, but
because any real measurement of a {\it single} dynamical variable or
the process of preparation of some state are not absolutely exact.
 It is easy to see from the
relation between the velocity $V^{i}=\frac{d\pi^i}{d\tau} $ in
$CP(N-1)$ and the energy variance $(\Delta H)^2$ through
Aharonov-Anandan relationship $\frac{dS}{d\tau}=\frac{2 \Delta H
}{\hbar}$ \cite{AA90}, where $\Delta H=\sqrt{<\hat H^2>-<\hat H>^2}$
is the uncertainty  of the Hamiltonian $\hat H$. Indeed, the quadric
form in the local coordinates is as follows: $dS^2 =G_{ik^*}d\pi^i
d\pi^{k*}=\frac{4(\Delta H)^2}{\hbar^2}d\tau^2$ and, therefore,
\begin{eqnarray}\label{uncer}
(\Delta H)^2 = \frac{\hbar^2}{4} G_{ik^*} \frac{d\pi^i}{d\tau} \frac
{d\pi^{k*}}{d\tau},
\end{eqnarray}
i.e. velocity $V^{i}$ in $CP(N-1)$ defines the variance of the
Hamiltonian.

But it is not the reason to deny a possibility to know any dynamical
variable with an acceptable accuracy.

\section{The geometric way of the gauge field generation}
The fundamental gauge field coming ``from nowhere" in the models of
elementary particles, and both Abelian \cite{Berry205} and
non-Abelian \cite{Wilczek} pseudo-potentials associated with
adiabatic Born-Oppenheimer approximation, have formally geometric
origins but of a different nature. Pseudo-potentials have a singular
source of monopole-like type whose nature arose under degeneration,
etc. But the mathematical artefact (singularity of mapping) cannon
be a reason for physical phenomenon. Dirac put the monopole as a
physical source of electromagnetic field. He assumed that
singularities are concentrated on some ``line of the knots" in the
physical space. However, the monopoles do not exist up to now as
physical object; one should have clear mechanism of the
non-integrability of the phase (or action functional).

The question is: is it possible to find a non-singular  fundamental
gauge potential if one uses not artificial parameter space, but an
inherently related projective Hilbert space $CP(N-1)$.

I argue that in the framework of my model the reason of anholonomy
is lurked in the curvature of the dynamical group manifold and its
invariant sub-manifold $CP(N-1)$. The ``fundamental'' interaction is
generated by the coset transformations \cite{Le2}. Their geometry is
the true source of some physical fields. It means that under the
affine parallel transport of LDV's agrees with the Fubini-Study
metric and there is some anholonomy depending on the curvature of
$CP(N-1)$.

If one uses the Berry formula (1.24) \cite{Berry198} in the local
coordinates $\pi^i$
\begin{eqnarray}
V_{ik*}(\pi^i)= \Im \sum_{a=0}^{N-1} \{\frac{\partial
g^{a*}}{\partial \pi^i} \frac{\partial g^a}{\partial \pi^{k*}} -
\frac{\partial g^{a*}}{\partial \pi^{k*}} \frac{\partial
g^a}{\partial \pi^i} \} =\Im \sum_{a=0}^{N-1}\{T^a_i T^{a*}_k -
(T^a_k)^* T^{a}_i \} \cr = -\Im [(1+ \sum |\pi^s|^2) \delta_{ik}-
\pi^{i^*} \pi^k](1+ \sum |\pi^s|^2)^{-2}= - \Im G_{ik^*}
\label{form},
\end{eqnarray}
one will find that it is closely related to Fubini-Study metric
(quantum metric tensor). There are two important differences between
original Berry's formula referring to arbitrary parameters and this
2-form in local coordinates inherently connected with eigen-problem.

1. The $V_{ik*}(\pi^i)$ is the singular-free expression.

2. It does not contain two eigen-values, say, $E_n, E_m$ explicitly
, but implicitly $V_{ik*}$ depends locally on the choice of single
$\lambda_p$ through the dependence in local coordinates
$\pi^i_{j(p)}$.

\section{Objective Quantum Measurement}
New scientific paradigm concerns the concept of the Muliverse or
omnium \cite{Penrose} where universal laws of Universe may be
contravened. I assume that this assumption is premature since it is
based on very speculative models of extended quantum objects. I
assume instead that our Universe has in fact infinite dimension
(even the state of a single hydrogen atom belongs to Hilbert space)
but quantum measurement reduce locally the Universe Geometry (UG)
down to 4D dynamical space-time whose geometry is a function of the
state of the measurement setup. It means that metric, connection,
curvature, etc. of the dynamical space-time (DST), are
microscopically state-dependent. Such reduction during objective
quantum measurement will be described below; this procedure leads to
different kind of the extended quantum objects - ``field shell" of
GCS.

Definitely, it is impossible  mathematically describe the state of
some real quantum setup. I will use a GCS
$(\pi^1_{j(p)},...,\pi^{N-1}_{j(p)})$ of some action operator
$\hat{S}=\hbar A(\hat{\eta^+}\hat{\eta})$ representing physically
distinguishable states. This means that any two points of $CP(N-1)$
define two ellipsoids differ at least by the orientations, if not by
the shape, as it was discussed above.

I assume that there is {\it expectation state} $|D>:
\hat{D}|D>=\lambda_p|D>$, associated with ``measuring device'' tuned
for measurement of dynamical variable $\hat{D}$ at some eigen-state
$(\pi^1_{j(p)},...,\pi^{N-1}_{j(p)})$
\begin{eqnarray}
|D>=|D_p(\pi^1_{j(p)},...,\pi^{N-1}_{j(p)})>
 =\sum_{a=0}^{N-1}
g^a(\pi^1_{j(p)},...,\pi^{N-1}_{j(p)})|\hbar a>=\sum_{a=0}^{N-1}
g^a|\hbar a>.
\end{eqnarray}
Hereafter I will omit indexes ${j(p)}$ for a simplicity. Now one
should build the spinor of the ``logical spin 1/2" in the local
basis $(|N>,|\widetilde{D}>)$ for the quantum question in respect
with the measurement of the local dynamical variable $\vec{D}$ at
corresponding GCS which may be marked by the local normal state
\begin{eqnarray}\label{45}
|N> = N^a|\hbar a>=(\frac{\partial^2 g^a}{\partial \pi^i \partial
\pi^k}-\Gamma_{ik}^s \frac{\partial g^a}{\partial \pi^s})
\frac{d\pi^i}{d\tau}\frac{d\pi^k}{d\tau}|\hbar a>.
\end{eqnarray}
Since in general $|D>$ it is not a tangent vector to $CP(N-1)$, the
deviation from GCS during the measurement of $\hat{D}$ will be
represented by tangent vector
\begin{eqnarray}\label{}
|\widetilde{D}>=|D>-<Norm|D>|Norm>=|D>-<N|D>\frac{|N>}{<N|N>}
\end{eqnarray}
defined as the covariant derivative on $CP(N-1)$. This operation is
the orthogonal projector $\hat{Q}$. Indeed,
\begin{eqnarray}
\widetilde{|\widetilde{D}>}= \widetilde{(|D>-<Norm|D>|Norm>)}\cr =
|D>-<Norm|D>|Norm> \cr - <Norm|(|D>-<Norm|D>|Norm>)|Norm> \cr
=|D>-<Norm|D>|Norm> = |\widetilde{D}>.
\end{eqnarray}
This projector $\hat{Q}$ takes the place of dichotomic dynamical
variable (quantum question) for the discrimination of the normal
state $|N>$ (it represents the eigen-state at GCS) the and the
orthogonal tangent state $|\widetilde{D}>$ that represents the
velocity of deviation form GCS. The coherent superposition of two
eigen-vectors of $\hat{Q}$ at the point $(\pi^1,...,\pi^{N-1})$
forms the spinor $\eta$ with the components
\begin{eqnarray}\label{513}
\alpha_{(\pi^1,...,\pi^{N-1})}=\frac{<N|D>}{<N|N>} \cr
\beta_{(\pi^1,...,\pi^{N-1})}=\frac{<\widetilde{D}|D>}
{<\widetilde{D}|\widetilde{D}>}.
\end{eqnarray}
Then from the infinitesimally close GCS
$(\pi^1+\delta^1,...,\pi^{N-1}+\delta^{N-1})$, whose shift is
induced by the interaction used for a measurement, one get a close
spinor $\eta+\delta \eta$ with the components
\begin{eqnarray}\label{514}
\alpha_{(\pi^1+\delta^1,...,\pi^{N-1}+\delta^{N-1})}=\frac{<N'|D>}
{<N'|N'>} \cr \beta_{(\pi^1+\delta^1,...,\pi^{N-1}+\delta^{N-1})}=
\frac{<\widetilde{D}'|D>}{<\widetilde{D}'|\widetilde{D}'>},
\end{eqnarray}
where the basis $(|N'>,|\widetilde{D}'>)$ is the lift of the
parallel transported $(|N>,|\widetilde{D}>)$ from the
infinitesimally close point
$(\pi^1+\delta^1,...,\pi^{N-1}+\delta^{N-1})$ back to the
$(\pi^1,...,\pi^{N-1})$. It is clear that such parallel transport
should be somehow connected with the variation of coefficients
$\Omega^{\alpha}$ in the dynamical space-time.

The covariance relative transition from one GCS to another
\begin{eqnarray}
(\pi^1_{j(p)},...,\pi^{N-1}_{j(p)}) \rightarrow
(\pi^1_{j'(q)},...,\pi^{N-1}_{j'(q)})
\end{eqnarray}
and the covariant differentiation (relative Fubini-Study metric) of
vector fields provides the objective character of the ``quantum
question" $\hat{Q}$ and, hence, the quantum measurement. This serves
as a base for the construction of the dynamical space-time as it
will be shown below.

These two infinitesimally close spinors may be expressed as
functions of $\theta,\phi,\psi,R$ and
$\theta+\epsilon_1,\phi+\epsilon_2,\psi+\epsilon_3,R+\epsilon_4,$
and represented as follows
\begin{eqnarray}\label{s1}
\eta = R \left( \begin {array}{c} \cos \frac{\theta}{2}(\cos
\frac{\phi- \psi}{2} - i\sin \frac{\phi - \psi}{2}) \cr \sin
\frac{\theta}{2} (\cos \frac{\phi+\psi}{2} +i \sin
 \frac{\phi+\psi}{2})  \end {array}
 \right)
 = R\left( \begin {array}{c} C(c-is) \cr S( c_1+is_1)
\end
{array} \right)
\end{eqnarray}
and
\begin{eqnarray}
&\eta+\delta \eta = R\left( \begin {array}{c} C(c-is) \cr S(
c_1+is_1) \end {array} \right) \cr + & R\left( \begin {array}{c}
S(is-c)\epsilon_1-C(s+i c)\epsilon_2+
C(s+ic)\epsilon_3+C(c-is)\frac{\epsilon_4}{R} \cr
 C(c_1+is_1)\epsilon_1+S(ic_1-s_1)\epsilon_2-S(s_1-ic_1)\epsilon_3
+S(c_1+is_1)\frac{\epsilon_4}{R}
\end
{array}
 \right)
\end{eqnarray}
may be connected with infinitesimal ``Lorentz spin transformations
matrix'' \cite{G}
\begin{eqnarray}
L=\left( \begin {array}{cc} 1-\frac{i}{2}\tau ( \omega_3+ia_3 )
&-\frac{i}{2}\tau ( \omega_1+ia_1 -i ( \omega_2+ia_2)) \cr
-\frac{i}{2}\tau
 ( \omega_1+ia_1+i ( \omega_2+ia_2))
 &1-\frac{i}{2}\tau( -\omega_3-ia_3)
\end {array} \right).
\end{eqnarray}
Then accelerations $a_1,a_2,a_3$ and angle velocities $\omega_1,
\omega_2, \omega_3$ may be found in the linear approximation from
the equation
\begin{eqnarray}\label{equ}
\eta+\delta \eta = L \eta
\end{eqnarray}
as functions of the ``logical spin 1/2" spinor components depending
on local coordinates $(\pi^1,...,\pi^{N-1})$.

Hence the infinitesimal Lorentz transformations define small
``space-time'' coordinates variations. It is convenient to take
Lorentz transformations in the following form $ct'=ct+(\vec{x}
\vec{a}) d\tau, \quad \vec{x'}=\vec{x}+ct\vec{a} d\tau
+(\vec{\omega} \times \vec{x}) d\tau$, where I put
$\vec{a}=(a_1/c,a_2/c,a_3/c), \quad
\vec{\omega}=(\omega_1,\omega_2,\omega_3)$ \cite{G} in order to have
for $\tau$ the physical dimension of time. The coordinates $x^\mu$
of points in this space-time serve in fact merely for the
parametrization of deformations of the ``field shell'' arising under
its motion according to non-linear field equations
\cite{Le5,Le6,Le7}.

\section{Field equations in the dynamical space-time}
The energetic packet  - ``particle'' associated with the ``field
shell'' is now described locally by the Hamiltonian vector field
$\vec{H}=\hbar
\Omega^{\alpha}\Phi^i_{\alpha}\frac{\partial}{\partial \pi^i} + c.c
$. Our aim is to find the wave equations for $\Omega^{\alpha}$ in
the dynamical space-time intrinsically connected with the objective
quantum measurement.

At each point $(\pi^1,...,\pi^{N-1})$ of the $CP(N-1)$ one has an
``expectation value'' of the $\vec{H}$ defined by a measuring
device. But a displaced GCS may by reached along one of the
continuum paths. Therefore the comparison of two vector fields and
their ``expectation values'' at neighboring points requires some
natural rule. The comparison makes sense only for the same
``particle'' or for its ``field shell'' along some path. For this
reason one should have an identification procedure. The affine
parallel transport in $CP(N-1)$ of vector fields is a natural and
the simplest rule for the comparison of corresponding ``field
shells''. Physically the identification of ``particle'' literally
means that its Hamiltonian vector field is a Fubini-Study covariant
constant.

Let me apply this parallel transport to the local Hamiltonian vector
field $H^i=\hbar \Omega^{\alpha}\Phi^i_{\alpha}$. The invariant
classification of the LDV's given above shows that $\vec{H}_b=\hbar
\Omega^b \Phi^i_b \frac{\partial}{\partial \pi^i}$ describes the
velocity of the shape variation of the ``ellipsoid of polarization",
but $\vec{H}_h=\hbar \Omega^h \Phi^i_h \frac{\partial}{\partial
\pi^i}$ gives the variation velocity of the orientation of this
ellipsoid in iso-space. Therefore  $\Omega^h$ may be treated as
``pure gauge fields" whereas $\Omega^b$ are ``matter field" whose
the mass spectrum $M_b=\hbar \Omega^b /c^2$ gives the axes of the
deformed ellipsoid.

Since we have only the unitary fields $\Omega^{\alpha}$ as
parameters of the GCS transformations we assume that in accordance
with an super-equivalence principle \cite{Le1,Le2} under the
infinitesimal variation of the unitary field $\delta
\Omega^{\alpha}$ in the dynamical space-time, the shifted
Hamiltonian field should coincide with the infinitesimal shift of
the tangent Hamiltonian field generated by the parallel transport in
$CP(N-1)$  during world time $\delta \tau$ \cite{H1}. Thus one has
\begin{equation}
\hbar (\Omega^{\alpha} + \delta \Omega^{\alpha} ) \Phi^k_{\alpha}=
\hbar \Omega^{\alpha}( \Phi^k_{\alpha} - \Gamma^k_{mn}
\Phi^m_{\alpha} V^n \delta \tau)
\end{equation}
and, hence, in accordance with the sufficiency criterion for the
equation with non-trivial solution
\begin{equation}
 \hbar(\delta \Omega^{\alpha} \delta^k_m+
 \Omega^{\alpha} \Gamma^k_{mn} V^n \delta \tau) \Phi^m_{\alpha}=0
\end{equation}
one has following equations
\begin{equation}
\frac{ \delta \Omega^{\alpha}}{\delta \tau} = -
\Omega^{\alpha}\Gamma^m_{mn} V^n,
\end{equation}
(there is no  summation in $m$).

We introduce the dynamical space-time coordinates $x^{\mu}$ as
state-dependent quantities, transforming in accordance with the
local Lorentz transformations $x^{\mu} + \delta x^{\mu} =
(\delta^{\mu}_{\nu} + \Lambda^{\mu}_{\nu} \delta \tau )x^{\nu}$. The
parameters of $\Lambda^{\mu}_{\nu} (\pi^1,...,\pi^{N-1})$ depend on
the local transformations of local reference frame in $CP(N-1)$
described in the previous paragraph. Then taking into account the
expressions for the ``4-velocity" $v^{\mu}=\frac{\delta
x^{\mu}}{\delta \tau} = \Lambda^{\mu}_{\nu} (\pi^1,...,\pi^{N-1})
x^{\nu} $ where
\begin{equation}
\matrix{ v^0&=&(\vec{x} \vec{a}) \cr
 \vec{v}&=&ct\vec{a}  +(\vec{\omega} \times
\vec{x}) \cr },
\end{equation}
one has the field equations
\begin{equation}\label{FSE}
v^{\mu} \frac{\partial \Omega^{\alpha}}{\partial x^{\mu} } = -
\Omega^{\alpha}\Gamma^m_{mn} V^n.
\end{equation}
One has in fact $N-1$ equations which may refer to the spectrum
generating fields $\Omega^b$. There are some questions concerning
this construction. First: the Dirac matrices $\gamma^{\mu}_{ss'}$
have the sense of some velocities, but the possible connection
between them and $v^{\mu}$ is not clear. The second question is: how
one can get $(N-1)^2$ equations for pure gauge fields $\Omega^h$?
Probably they relate to Killing fields in $CP(N-1)$, but I have not
now answers on these questions.

\vskip 0.2cm CONCLUSION

1. It is proposed the generalized (in comparison with ``2-level"
case \cite{Le1}) the intrinsically geometric scheme of the quantum
measurement of an arbitrary Hermitian ``N-level" dynamical variable.
The interaction arose due to the breakdown of $G=SU(N)$ symmetry is
used for such measurement and it is represented by the affine gauge
``field shell" propagated in the dynamical state-dependent
space-time.

2. The concept of ``super-relativity" \cite{Le2} is in fact a
different kind of attempts of ``hybridization" of the internal and
space-time symmetries. In distinguish from SUSY where a priori
exists the ``extended space-time - super-space", in my approach the
dynamical space-time arises only as a parametrization of $SU(N)$
local dynamical variables under ``yes/no" quantum measurement.

3. The pure local formulation of theory in $CP(N-1)$ leads seemingly
to the decoherence \cite{Le1}. We may, of course, to make mentally
the concatenation of any two quantum systems living in direct
product of their state spaces. The variation of the one of them
during a measurement may lead formally to some variations in the
second one. Unavoidable fluctuations in our devices may even confirm
predictable correlations. But the introduction of the
state-dependent dynamical space-time evokes a necessity to
reformulate the Bell's inequalities which may lead then to a
different condition for the coincidences.

4. The locality in the quantum phase space $CP(N-1)$ leads to
extended quantum particles - ``field shell" that obey the
quasi-linear PDE \cite{Le5,Le6,Le7}. The physical status of their
solutions is the open question. But if they somehow really connected
with ``elementary particles", say, electrons, then the plane waves
of de Broglie should not be literally refer to the state vector of
the electron itself but rather to covector (1-form) realized, say,
as a periodic cristall lattice. The fact that the condition for
diffraction is in nice agreement with experiments may be explained
that for this agreement it is important only {\it relative velocity}
of electron and the lattice. The fact that momentum of electron
$p_k=\frac{\partial W}{\partial q^k} $ and the group velocity
$v^k=\frac{dq^k}{dt}$ should be in opposite directions may be easy
proved. Schr\"odinger \cite{Schr4} pointed out that the
Hamilton-Jacobi equation for the material point with the unit mass
$m=1$
\begin{equation}
\frac{\partial W}{\partial t}+T(q^k,\frac{\partial W}{\partial
q^k})+V(q^k)=0
\end{equation}
may be understood as describing the surfaces of constant phase of
the action waves. If
\begin{equation}
 W=-Et+S(q^k),
 \end{equation}
then one gets
\begin{equation}
T(q^k,\frac{\partial W}{\partial q^k})=E-V(q^k).
\end{equation}
The Hertz metric
\begin{equation}
dS^2=2 T(q^k,\frac{dq^k}{dt})dt^2=2(E-V(q^k))dt^2=g_{ik}dq^i dq^k.
\end{equation}
may be interpreted as a some effective ``index of refraction"
\begin{equation}
n=\sqrt{2 T} = \sqrt{g_{ik}v^i v^k}=\sqrt{2(\hbar \omega -V)}
\end{equation}
for the action waves with the positive phase velocity
\begin{equation}
u=\frac{E}{\sqrt{2 T}} =\frac{E}{ \sqrt{g_{ik}v^i v^k}}=\frac{\hbar
\omega}{\sqrt{2(\hbar \omega -V)}}.
\end{equation}
But the scalar product
\begin{equation}
p_k v^k=\frac{\partial W}{\partial
q^k}\frac{dq^k}{dt}=\frac{\partial W}{\partial t}=-E
\end{equation}
shows that the momentum and the group velocity should be in opposite
directions.

\vskip 0.2cm ACKNOWLEDGEMENTS

I am sincerely grateful Larry Horwitz for interesting discussions
and critical notes.


\vskip 0.2cm
\begin{thebibliography}{99}
\bibitem{Penrose}
R.Penrose, {\it The Road to Reality}, Alfred A.Knopf, New-York,
(2005).
\bibitem{Heg}
G.C.Hegerfeldt, Phys.Rev.D{\bf 10}, 3320 (1974).
\bibitem{Ash}
A.Ashtekar and T.A.Schilling, arXive: gr-qc/9706069.
\bibitem{Le1}
P.Leifer, JETP Letters, {\bf 80}, (5) 367 (2004).
\bibitem{Schr1}
E.Schr\"odinger, Brit.J.Phil.Sci.,{\bf 3}, 233 (1952).
\bibitem{Schr2}
E.Schr\"odinger, Nuovo Cimento.,{\bf 1}, 5 (1955).
\bibitem{Feynman}
R.P.Feynman, {\it QED - The strange theory of light and matter},
Princeton University Press, New Jersey, (1985).
\bibitem{Schr3}
E.Schr\"odinger, Naturwissenschaften, {\bf 14}, H.28 (1926).
\bibitem{Schr4}
E.Schr\"odinger, Ann. Physik.,{\bf 79}, 361 (1926).
\bibitem{Gelfand}
I.M.Gelfand, {\it Lectures on Linear Algebra}, ``Nauka", Moscow,
(1966).
\bibitem{Berry198}
M.V.Berry, {\it Lectures on Quantum Adiabatic Anholonomy}, (1990).
\bibitem{Dirac1}
P.A.M. Dirac, {\it Lectures on quantum field theory}, New York,
Yeshiva University, 1967.
\bibitem{Le2}
P.Leifer, Found. Phys. {\bf 27}, (2) 261 (1997).
\bibitem{Le3}
P.Leifer, arXive: physics/0601201.
\bibitem{Onf}
F.P.Onufrieva, JETP, {\bf 89}, 6(12) 2270 (1985).
\bibitem{Ost}
V.S.Ostrovskii, JETP, {\bf 91}, 5(11) 2270 (1986).
\bibitem{Le4}
P.Leifer, {\it Dynamics of the spin coherent states}, Ph.D Thises,
Physical institute of low temperatures for technical and
engineering, Kharkov, 1990.
\bibitem{H1}
L.P.Horwitz, arXive: hep-ph/9606330.
\bibitem{Le5}
P.Leifer, arXive: gr-qc/0503083.
\bibitem{Le6}
P.Leifer and L.P.Horwitz, arXive: gr-qc/0505051, V.2.
\bibitem{AA90}
J.Anandan and Y.Aharonov, Phys.Rev.Lett., {\bf 65}, 1697 (1990).
\bibitem{Berry205}
M.V.Berry and R.Lim, J.Phys. A: Nath.Gen.,{\bf 23}, 655 (1990).
\bibitem{Wilczek}
F.Wilczek and A.Zee, Phys.Rev.Lett., {\bf 52}, (24), 2111 (1984).
\bibitem{G}
C.W.Misner, K.S.Thorne, J.A.Wheeler,{\it Gravitation}, W.H.Freeman
and Company, San Francisco, 1973.
\bibitem{Le7}
P.Leifer, Found.Phys.Lett., {\bf 18}, (2) 195 (2005).

\end{thebibliography}
\end{document}